\documentstyle[preprint,aps,prl]{revtex}
\begin{document}
\draft
\title{First-principle study of excitonic self-trapping in diamond}

\author{Francesco Mauri \cite{Address} and Roberto Car}
\address{
Institut Romand de Recherche Num\'erique en Physique
des Mat\'eriaux (IRRMA)  IN-Ecublens 1015 Lausanne, Switzerland}
%\date{Received}
\maketitle
\begin{abstract}
We present a first-principles study of excitonic self-trapping
in diamond. Our calculation provides evidence for self-trapping
of the $1s$ core exciton and gives a coherent
interpretation of recent experimental X-ray absorption and
emission data.
Self-trapping does not occur in the case of a single valence
exciton. We predict, however, that self-trapping should occur
in the case of a valence biexciton. This process is accompanied by
a large local relaxation of the lattice which could be observed
experimentally.
\end{abstract}
Diamond presents an unusually favorable combination of characteristics
that, in connection with the recent development of techniques
for the deposition of thin diamond films, make this material a good candidate
for many technological applications.
Particularly appealing is the use of diamond in electronic or in
opto-electronic devices, as e.g.  UV-light emitting devices.
Moreover, diamond is an ideal material for the construction of windows that
operate under high power laser radiation or/and in adverse environments.
It is therefore interesting to study
radiation induced defects with deep electronic levels in the gap, since
these can have important implications in many of these applications.

Excitonic self-trapping is a possible mechanism for the formation
of deep levels in the gap. The study of such processes in
a purely covalent material, like diamond,
is interesting also from a fundamental point of view.
Indeed, excitonic self-trapping has been studied so far
mostly in the context of ionic compounds, where it is always
associated with, and often driven by, charge transfer effects.
In a covalent material the driving mechanism for self-trapping is instead
related to the difference in the bonding character between the valence and
the conduction band states.

Both experimental data and theoretical arguments
suggest the occurrence of self-trapping processes in diamond.
In particular, a nitrogen (N) substitutional impurity induces
a strong local deformation of the lattice
\cite{NitrogenEPR,NitrogenDRT,Nitrogentheo} that can
be interpreted as a self-trapping of the donor electron.
The structure of a $1s$ core exciton is more controversial
\cite{M84,JP91,N92,B93,MSWGMGN,MS94}. Indeed
the similarity between an excited core of carbon and
a ground-state core of nitrogen suggests that the core exciton should
behave like a N impurity. However,
the position of the core exciton peak in the diamond K-edge
absorption spectra is only 0.2 eV lower than the conduction
band minimum \cite{M84,B93,MSWGMGN}, while
a N impurity originates a deep
level 1.7 eV below the conduction band edge\cite{Nitrogenlevpos}.
On the other hand, emission spectra \cite{MSWGMGN}
suggest that a $1s$ core exciton should self-trap like a N impurity.
Finally, we consider valence excitations. In this case
experimental evidence indicates that
a single valence exciton is of the Wannier type, i.e. there is no
self-trapping.
To our knowledge, neither experimental nor theoretical
investigations on the behavior of a valence
biexciton in diamond have been performed, although simple scaling
arguments suggest that
the tendency to self-trap should be stronger for biexcitons
than for single excitons.

In this letter, we present a detailed theoretical
study of excitonic self-trapping effects in diamond.
In particular, we have investigated the
Born-Oppenheimer (BO) potential energy surfaces corresponding
to a core exciton, a valence exciton and a valence
biexciton in the context of density functional theory (DFT),
within the local density approximation (LDA) for exchange and correlation.
Our calculation indicates that the $1s$ core exciton is on a different BO
surface in absorption and in emission experiments.
Indeed X-ray absorption creates excitons
in a $p$-like state as required by dipole selection rules.
Subsequently the system makes a transition to an $s$-like state
associated to a self-trapping distortion of the atomic lattice,
similar to that found in the N impurity case. These results provide
a coherent interpretation of the experimental data.
In addition, our calculation suggests that self-trapping should also
occur for a valence biexciton.
This is a prediction that could be verified experimentally.

Let us start by discussing a simple model \cite{HS,MC2}.
In diamond, the occupied valence and the lower
conduction band states derive from superpositions of atomic $sp^3$
hybrids having bonding and antibonding character, respectively.
Thus, when an electron, or a hole, or an electron-hole pair is added
to the system, this can gain in deformation energy
by relaxing the atomic lattice. Scaling arguments suggest that
the deformation energy gain
$E_{ def}\propto - 1/N_{ b}$, where $N_{b}$ is the number of bonds over
which the perturbation is localized. This localization, in turn, has
a kinetic energy cost $E_{ kin}\propto + 1/N_{ b}^{2/3}$ due to quantum
confinement.
The behavior of the system is then governed by the value of $N_{ b}$
that minimizes the total energy $E_{ sum}=E_{ def}+E_{ kin}$.
Since the only stationary point of $E_{ sum}$ is a maximum, $E_{ sum}$
attains its minimum value at either one of the two extrema $N_{ b}=1$
or $N_{ b}=\infty$. If the minimum occurs for
$N_{ b}=1$, the perturbation is self-trapped
on a single bond which is therefore stretched.
If the minimum occurs for $N_{ b}=\infty$, there is no self-trapping and
the perturbation is delocalized.
When $N_{p}$ particles (quasi-particles)
are added to the system, one can show that,
for a given $N_{ b}$, $E_{ def}$ scales as $N_{ p}^2$,
while $E_{ kin}$ scales as $N_{ p}$. As a consequence,
the probability of self-trapping is enhanced when $N_{p}$ is larger.
This suggests that biexcitons should have
a stronger tendency to self-trap than single
excitons\cite{MC2,nomorethan2}.

In order to get a more quantitative understanding of self-trapping
phenomena in diamond, we
performed self-consistent electronic structure calculations,
using norm-conserving pseudopotentials \cite{bhs82} to describe
core-valence interactions.
The wave-functions and the electronic density were expanded in
plane-waves with a cutoff of 35 and of 140 Ry, respectively.
We used a periodically repeated simple cubic supercell containing 64
atoms at the experimental equilibrium lattice constant.
Only the wave-functions at the $\Gamma$ point were considered.
Since the self-trapped states are almost completely localized on one
bond, they are only weakly affected by the boundary conditions in a
64 atom supercell. The effect of the $k$-point sampling was analysed
in Ref.~\cite{Nitrogentheo} where similar calculations for a N impurity
were performed using the same supercell. It was found that
a more accurate $k$-point sampling
does not change the qualitative physics of the distortion but only
increases the self-trapping energy by 20$\%$ compared to
calculations based on the $\Gamma$-point only\cite{Nitrogentheo}.

In order to describe a core exciton we adopted the method
of Ref.~\cite{PS93}, i.e. we generated a norm conserving
pseudopotential for an excited carbon atom with one electron in the
$1s$ core level
and five electrons in the valence $2s$-$2p$ levels.
In our calculations for a valence exciton or biexciton
we promoted one or two electrons, respectively, from the highest valence
band state to the lowest conduction band state.
Clearly, our single-particle approach cannot account for the (small)
binding energy of delocalized Wannier excitons. However
our approach should account for the most important contribution
to the binding energy in the case of localized excitations.
Structural relaxation studies were based on the Car-Parrinello (CP)
approach\cite{CP85}. We used a standard CP scheme for both the core and
the valence exciton, while a modified CP dynamics, in which
the electrons are forced to stay in an arbitrary excited
eigenstate\cite{MC2,MCT93}, was necessary
to study the BO surfaces corresponding to a valence biexciton.
All the calculations were made more efficient by
the acceleration methods of Ref.~\cite{TMC94}.

We first computed the electronic structure
of the core exciton with the atoms in
the ideal lattice positions.
In this case the excited-core atom induces two defect states in the
gap: a non-degenerate level
belonging to the A$_1$ representation of the T$_{\rm d}$ point group,
0.4 eV below the conduction band edge,
and a 3-fold degenerate level with T$_2$ character,
0.2 eV  below the conduction band edge.
By letting the atomic coordinates free to relax, we found that the absolute
minimum of the A$_1$ potential energy surface correponds
to an asymmetric self-trapping distortion of the lattice
similar to that found for
the N impurity\cite{Nitrogentheo}.
In particular, the excited-core atom and its nearest-neighbor, labeled
{\it a} and {\it b}, respectively, in Fig.~\ref{IS}, move away from
each other on the (111) direction.
The corresponding displacements from the ideal sites are equal to
10.4$\%$ and to 11.5$\%$
of the bond length, respectively, so that the {\it (a,b)}-bond
is stretched by 21.9$\%$.
The other atoms move very little: for instance the nearest-neighbor
atoms labeled {\it c} move by 2.4$\%$ of the bond length only.
This strong localization of the distortion is
consistent with the simple scaling arguments discussed above.

As a consequence of the atomic relaxation, the non-degenerate level
ends up in the gap at 1.5 eV below the conduction band edge, while
the corresponding wavefunction localizes on the stretched bond.
The 3-fold degenerate level remains close to the
conduction band edge, but since the distortion lowers the symmetry from
T$_{\rm d}$ to C$_{\rm 3v}$, the 3-fold degenerate level splits
into a 2-fold degenerate E level and a non-degenerate A$_1$ level.

In Fig.~\ref{CDCE} we report the behavior of the potential energy
surfaces corresponding to the ground-state, the A$_1$ and the T$_2$
core exciton states as a function of the self-trapping distortion.
Notice that the distortion gives a total energy gain of 0.43 eV
on the A$_1$ potential energy surface.
The same distortion causes an increase of the ground-state energy
of 1.29 eV.

Our calculation indicates that the core-exciton behaves like
the N impurity\cite{Nitrogentheo}, supporting, at least qualitatively,
the validity of the equivalent core approximation.
The similar behavior of the A$_1$ level in the core exciton
and in the N impurity case was also pointed out recently
in the context of semi-empirical
CNDO calculations\cite{MS94}. The differences between the core exciton
and the impurity\cite{Nitrogentheo}
are only quantitative: in particular, the relaxation
energy
and especially the distance of the A$_1$ level from the
conduction band edge are smaller for the core exciton than
for the N impurity.

Our results suggest the following interpretation of the
experimental data of Refs.~\cite{M84,MSWGMGN}:
(i) During X-ray absorption the atoms are in the ideal lattice positions.
Dipole transitions from a $1s$ core level to a A$_1$ valence level
are forbidden, but transitions to the T$_2$ level are allowed.
In our calculation the T$_2$ level is 0.2 eV lower than the conduction
band edge, in good agreement with the core exciton peak
observed in X-ray absorption spectra \cite{M84,MSWGMGN}.
(ii) On the T$_2$ BO potential energy surface the lattice undergoes a
Jahn-Teller distortion which lowers its energy (see Fig.~\ref{CDCE}).
(iii) Since the LO phonon energy in diamond (0.16 eV) is comparable to the
energy spacing between the A$_1$ and the T$_2$ surfaces, which is less than
0.2 eV after the Jahn-Teller distortion, the probability of a non-adiabatic
transition from the T$_2$ to the A$_1$ surface is large.
(iv) On the A$_1$ level the system undergoes a strong lattice relaxation
resulting in a localization of the exciton on a single bond.
(v) The self-trapping distortion induces a Stokes shift in the
emitted photon energy. If the atomic relaxation were complete the
Stokes shift would be equal to 1.9 eV, which correponds
(see Fig.~\ref{CDCE}) to the energy dissipated in the T$_2$-A$_1$
transition (0.2 eV), plus the energy gained by
self trapping on the A$_1$ surface (0.43 eV), plus the energy
cost of the self-trapping distortion
on the ground-state energy surface (1.29 eV).

The data reported in Ref.\cite{MSWGMGN} show a
shift of about 1 eV in the positions
of the peaks associated to the $1s$ core exciton
in X-ray absorption and emission spectra. The emission peak
is very broad, with a large sideband that corresponds to Stokes
shifts of up to 5 eV.
As pointed out in Ref.~\cite{MSWGMGN}, this large sideband
is likely to be the effect of incomplete relaxation. This
is to be expected since
the core exciton lifetime should be comparable to the
phonon period\cite{MSWGMGN}. As a consequence, the atomic lattice
would be able to perform only a few damped oscillations
around the distorted minimum structure during the lifetime of
the core exciton.

We now present our results for the valence excitations.
While in the case of a single exciton the energy is minimum for the
undistorted crystalline lattice, in the case of a biexciton we find that
the energy is minimized in correspondence of a localized distortion
of the atomic lattice. This is characterized by a large outward symmetric
displacement along the (111) direction of the atoms
{\it a} and {\it b} in Fig.~\ref{IS}. As a result the
{\it (a,b)}-bond is broken since the distance between the atoms
{\it a} and {\it b} is increased by 51.2$\%$ compared to the
crystalline bondlength.
This distortion can be viewed as a kind of local graphitization
in which the atoms {\it a} and {\it b} change from fourfold
to threefold coordination and the corresponding hybridized orbitals
change from $sp^3$ to $sp^2$ character.
Again, in agreement with the model based on simple scaling arguments,
the distortion is strongly localized on a single bond.
As a matter of fact and with reference to the Fig.~\ref{IS},
the atoms {\it c} and {\it d} move by 1.2$\%$ of the bondlength, the
atoms {\it e} and {\it f} move by 2.3$\%$,
and the atoms not shown in the figure by less than 0.9$\%$.

The self-trapping distortion of the biexciton gives rise to
two deep levels in the gap:
a doubly occupied antibonding level, at 1.7 eV below
the conduction band edge, and an empty bonding level,
at 1.6 eV above the valence band edge. Both states are localized on the
broken bond.

In Fig.~\ref{CDVE} we show how different BO potential energy surfaces
behave as a function of the self-trapping distortion of the valence biexciton.
In particular, from this figure we see that, while for the biexciton there
is an energy gain of 1.74 eV  in correspondence with the self-trapping
distortion, the same distortion has an energy cost of 1.49 eV for
the single exciton, and of 4.85 eV for the unexcited crystal.
We notice that, while DFT-LDA predicts self-trapping
for the valence biexciton,
it does not do so for the single exciton, in agreement with experiment.

Similarly to the case
of the core exciton the major experimental
consequence of the self-trapping of the
valence biexciton is a large Stokes shift in the
stimulated-absorption spontaneous-emission
cycle between the exciton and the biexciton BO surfaces.
As it can be seen from Fig.~\ref{CDVE},
this Stokes shift should be equal to 3.23 eV, i.e. to
the sum of the energy gain
of the biexciton (1.74 eV) and of the energy cost of the exciton
(1.49 eV) for the self-trapping relaxation.
The fundamental gap of diamond is indirect. Thus the spontaneous decay
of a Wannier exciton in an ideal diamond crystal is phonon assisted and
the radiative lifetime of the exciton is much longer than in direct
gap semiconductors.
However, after self-trapping of the biexciton,
the translational symmetry is broken
and direct spontaneous emission becomes allowed.
As a consequence the radiative life time of the
self-trapped biexciton is much smaller than that of the Wannier exciton.
Using the DFT-LDA wavefunctions,
we obtained a value of $\sim 7$ ns for the
radiative lifetime of the biexciton within the dipole approximation.
This is
several orders of magnitude larger than the typical phonon period.
Therefore the self-trapping relaxation of
the valence biexciton should be completed before the radiative decay.

A self-trapped biexciton is a bound state of two excitons strongly localized
on a single bond. Thus
the formation of self-trapped biexcitons requires
a high excitonic density.
To realize this condition it is possible either to excite directly bound
states of Wannier excitons,
or to create a high density electron-hole plasma, e.g. by strong laser
irradiation. In the second case many self-trapped biexcitons could be
produced. This raises some interesting implications.
If many self-trapped biexcitons are created, they
could cluster producing a macroscopic graphitization.
Moreover, since the process of self-trapping is associated
with a relevant energy
transfer from the electronic to the ionic degrees of freedom,
in a high density electron hole plasma biexcitonic
self-trapping could heat the crystal up to the melting point
in fractions of a ps, i.e in the characteristic time of
ionic relaxation. Interestingly, melting of a GaAs crystal
under high laser irradiation has been observed to occur in fractions of a
ps \cite{GaAs91}.
In Ref.~\cite{GaAs91} this phenomenon has been ascribed to the change
in the binding properties due to the electronic excitations.
Our study on diamond leads one to speculate
that in a sub-picosecond melting experiment
self-trapping phenomena could play an important role.

In conclusion, we have studied
excited-state BO potential energy surfaces of
crystalline diamond within DFT-LDA.
Our calculation predicts
self-trapping of the core exciton and provides
a coherent description of the X-ray
absorption and emission processes, which
compares well with the experimental data.
Moreover, we also predict self-trapping of the valence biexciton, a
process characterized by a large local lattice relaxation.
This implies a strong Stokes shift in the
stimulated absorption - spontaneous emission cycle of about 3 eV,
which could be observed experimentally.

It is a pleasure to thank F.~Tassone
for many useful discussions.
We acknowledge support from the Swiss National Science Foundation under
grant No.~20-39528.93

\begin{figure}
\caption{Atoms and bonds in the ideal diamond crystal (left panel).
Atoms and bonds after the self-trapping distortion associated
with the valence biexciton (right panel). In this case the distance between
the atoms {\it a} and {\it b} increases by 51.2$\%$. A similar but
smaller distortion is associated with the core exciton:
in this case the {\it (a,b)} distance is increased by 21.9$\%$.
\label{IS}}
\end{figure}
\begin{figure}
\caption{Total energy vs self-trapping distortion
of the core-exciton. The figure displays the
BO potential energy surfaces correponding to
the ground-state, the A$_1$, and the T$_2$ core exciton states.
\label{CDCE}}
\end{figure}
\begin{figure}
\caption{
Total energy as a function of the self-trapping distortion
of the biexciton. The BO energy surfaces correponding to
the ground state, the valence exciton, and the
valence biexciton are shown in the figure.
\label{CDVE}}
\end{figure}


\begin{references}
\bibitem{nomorethan2}
The number of equal particles that can be accommodated on one
bond of the
crystal in the same quantum state is limited by the Pauli
principle.
Thus no more than two holes or/and two electrons with opposite
spins can be localized on one bond of a $sp^3$ bonded semiconductor.
\end{references}
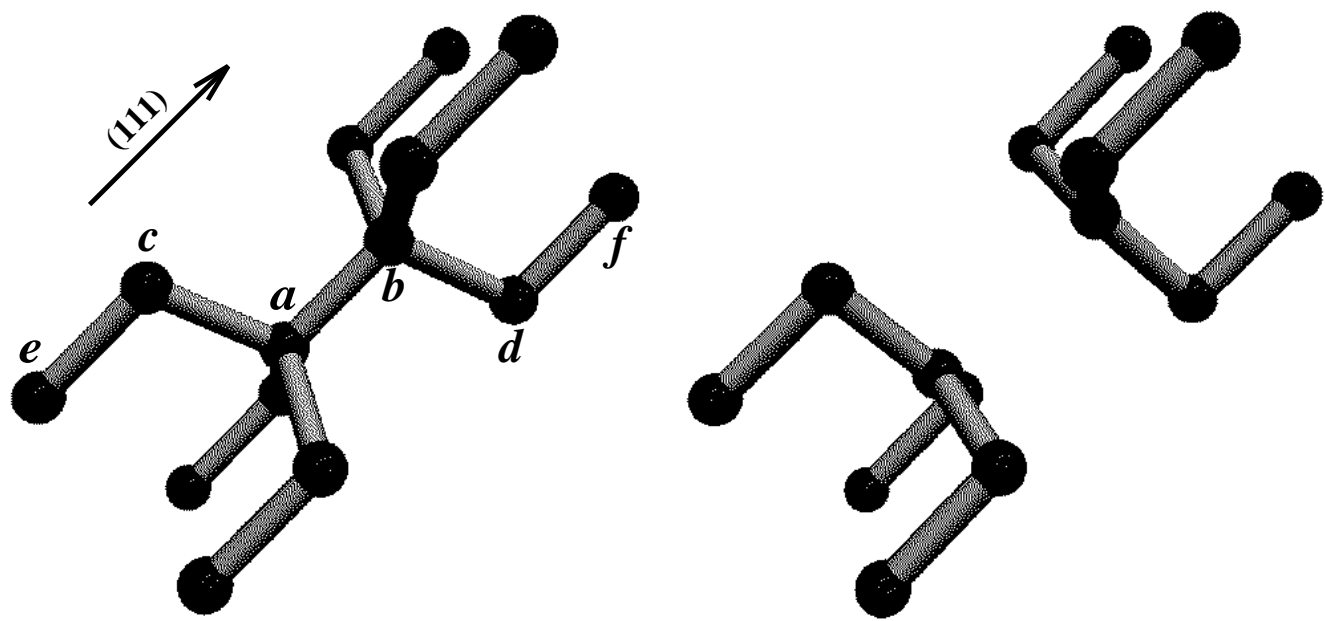
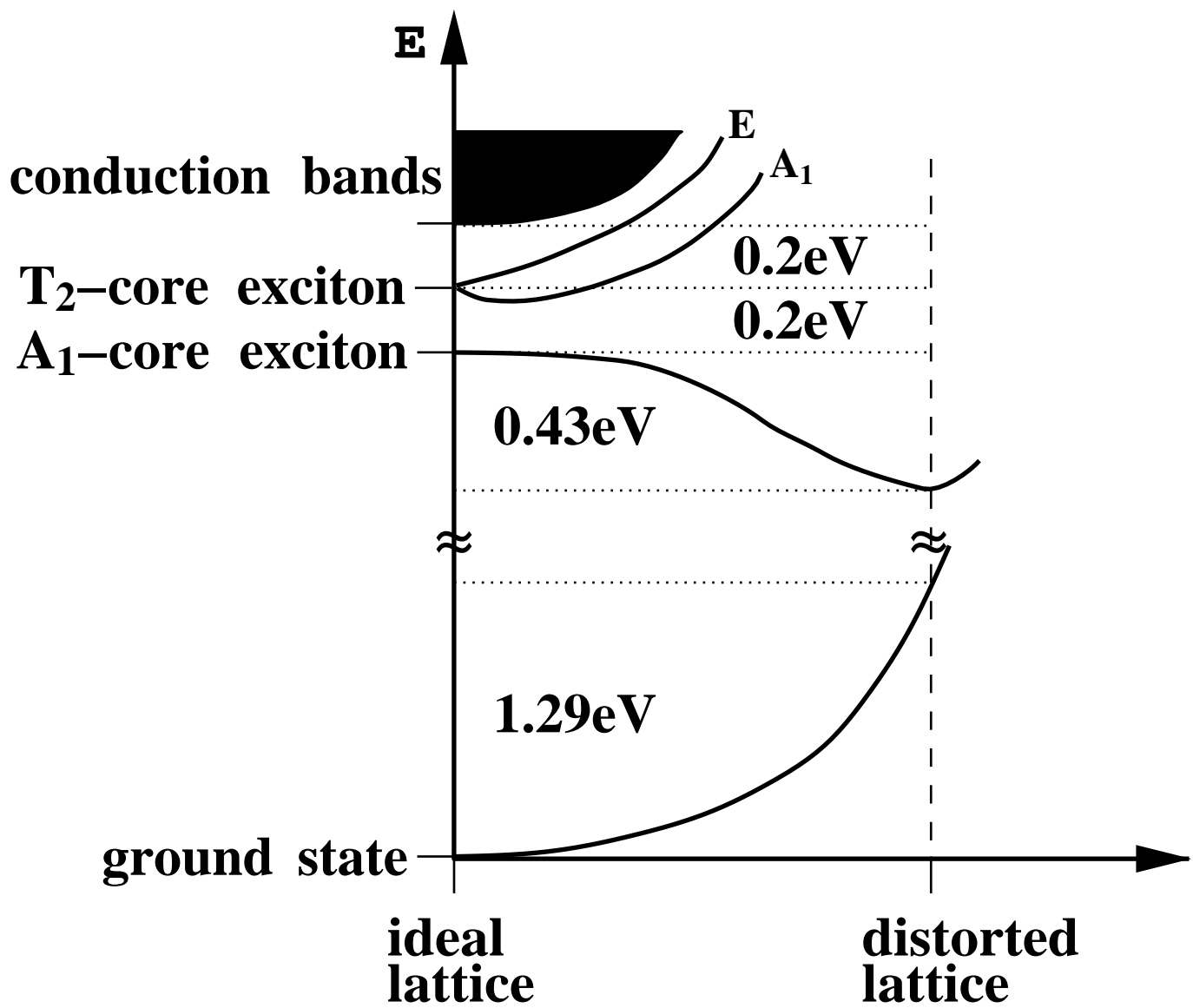
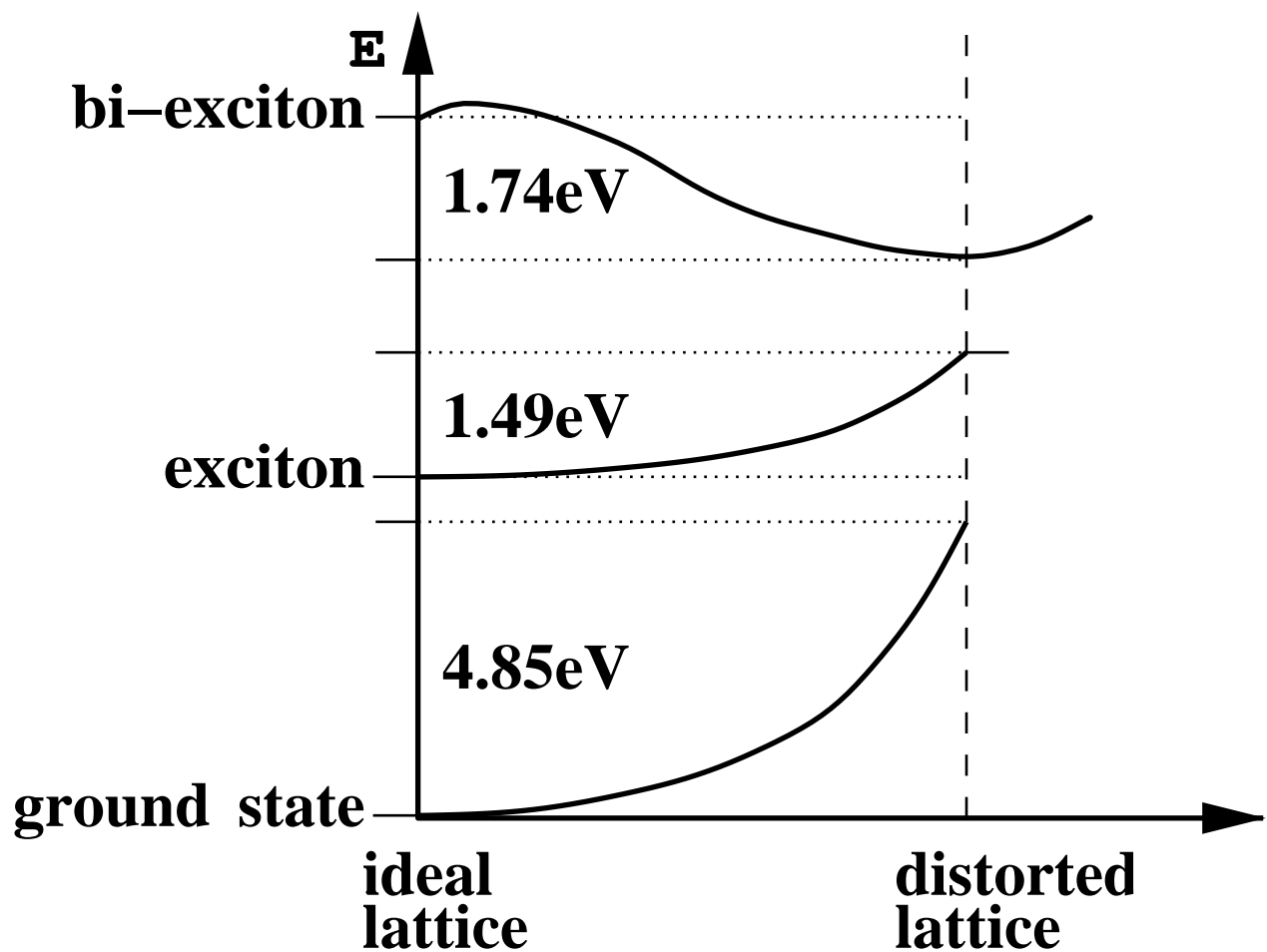
\end{document}